# Detection Probability of Terrestrial Radio Signals by a Hostile Super-civilization

Dr. Alexander L. Zaitsev, IRE, Russia, alzaitsev@gmail.com

Comparison of the total number of the radar astronomy transmissions with respect to that used for sending messages to extra-terrestrial civilizations reveals that the probability of detection of the radio signals to extraterrestrials (ETs) is one million times smaller than that of the radar signals used to study planets and asteroids in the Solar System.

There are three large-dish instruments in the world that are currently employed for doing radar investigations of planets, asteroids and comets [1]: ART (Arecibo Radar Telescope), GSSR (Goldstone Solar System Radar), and EPR (Evpatoria Planetary Radar). Radiating power and directional diagram of these instruments is so outstanding that it also allows us to emit radio messages to outer space, which are practically detectable everywhere in the Milky Way. This dedicated program is called METI (Messaging to Extra-Terrestrial Intelligence) [2], as contrasted to SETI (Search for Extra-Terrestrial Intelligence), [3].

Recently, some scientists and SF writers have expressed their concern [4] that sending messages to the stars in our galaxy, which may have a habitable life, jeopardizes existence of our own civilization because our signals helps ETs to pin down location of the Solar System in the Milky Way. If the Aliens reached the level of a super-civilization, it might send a space fleet to the Earth to either destroy it or to convert us to slaves.

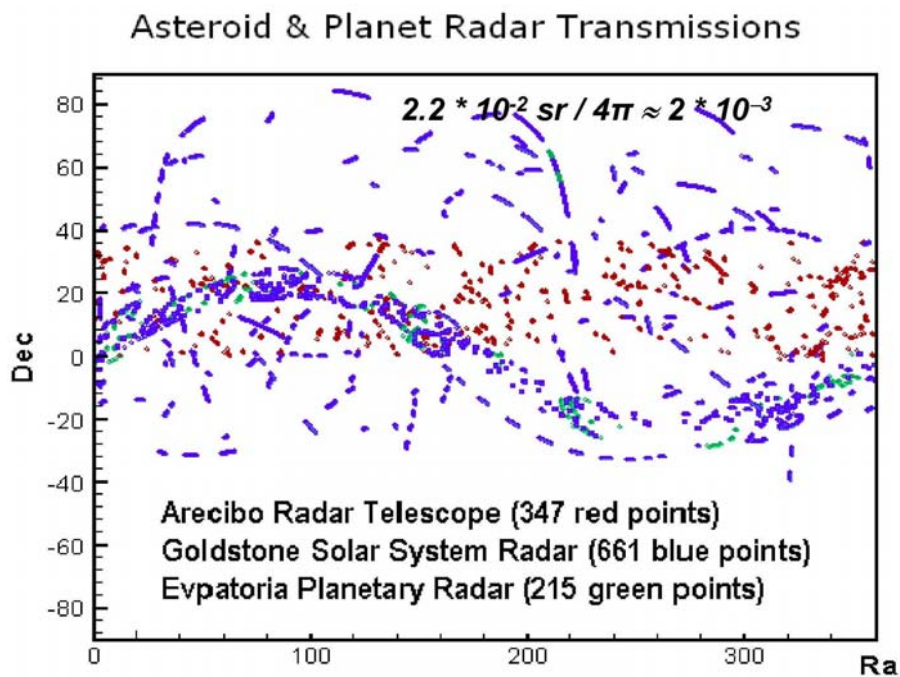

The goal of this letter is to estimate the probability of detection of the terrestrial radio signals by a presumable hostile super-civilization existing somewhere in our galaxy. Our calculation starts from the notice that over all the radar astronomy history about 1,400 sets of radio transmissions were produced. Their distribution all over the sky is shown in Figure 1 in the plane of ecliptic coordinates.

The total area of the sky illuminated by these transmissions, is about 0.022 steradians (sr), or about $2·10^{-3}$ part of the whole sky. The total number of METI transmissions is 16 sets only [5], and the total area of sky, illuminated by the METI transmissions, is about $10^{-5}$ sr, or 2000 times less than that covered by radar astronomy transmissions, Figure 2.

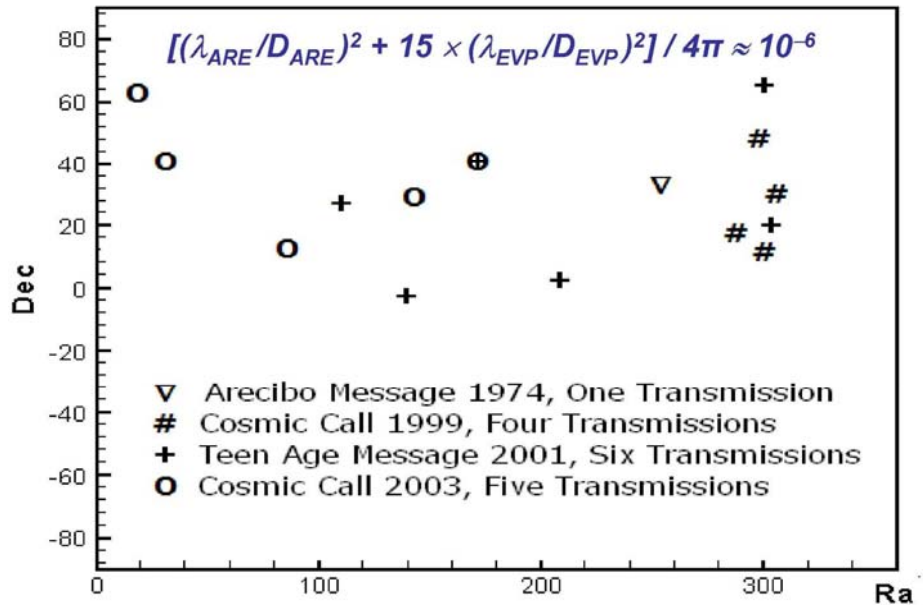

Total duration of time of radar transmissions exceeds the overall time interval of the METI transmissions by a factor of 500. Therefore, we can conclude that the probability to detect the radar astronomy transmissions by a hostile super-civilization is $2000 \times 500 = 1,000,000$ times higher than that of the METI transmissions.

So, if someone is concerned about our detection by an aggressive super-civilization (so-called *METI-phobia,* [6]), first of all one has to prohibit not the METI, but the radar astronomy. However, one can not prohibit it because the radar astronomy is an important and indispensable component of the asteroid hazard and defense system [7]. For this reason, we conclude that all on-going conversations about the ETs danger of METI for our civilization are meaningless, and the radar astronomy instruments should remain open for doing further exploration of the interstellar space with METI transmissions.